\documentclass[a4paper,fleqn]{cas-dc}

\usepackage[numbers]{natbib}

\def\tsc#1{\csdef{#1}{\textsc{\lowercase{#1}}\xspace}}
\tsc{WGM}
\tsc{QE}
\tsc{EP}
\tsc{PMS}
\tsc{BEC}
\tsc{DE}

\begin{document}
\let\WriteBookmarks\relax
\def\floatpagepagefraction{1}
\def\textpagefraction{.001}
\shorttitle{Theoretical evidence of a significant modification of the electronic structure of DWNTs due to the interlayer interaction}
\shortauthors{VN Popov}

\title [mode = title]{Theoretical evidence of a significant modification of the electronic structure of double-walled carbon nanotubes due to the interlayer interaction}                      

\author[1]{Valentin N. Popov}

\cormark[1]


\address[1]{Faculty of Physics, University of Sofia, BG-1164 Sofia, Bulgaria}

\cortext[cor1]{Corresponding author}

\begin{abstract}
The recently reported experimental optical spectra of double-walled carbon nanotubes exhibit more peaks than it could be expected based on the layers alone. The appearance of excess peaks has been attributed to the interlayer interaction. In order to elucidate the origin of the excess peaks, we perform calculations of the optical absorption of a particular nanotube using the recursion method with non-orthogonal tight-binding basis functions. Our study shows that the interlayer interaction can give rise to major changes in the electronic structure of this nanotube, manifesting themselves with shifts of the optical transitions and appearance of new optical transitions. The derived absorption spectrum is found to be in excellent agreement with the available experimental data, which justifies the use of the proposed approach for double-walled carbon nanotubes.  
\end{abstract}



\maketitle

\section{Introduction}

The double-walled carbon nanotubes (DWNTs) consist of two coaxial cylindrical graphitic layers, interacting with each other by weak Van der Waals interactions. These structures have attracted much attention because they are ideal systems to study the influence of the interlayer interaction on the physical properties \cite{pfei08}, as well as because of their future application \cite{shen11}. The characterization of the DWNTs is usually performed by means of high-precision experimental techniques including spectroscopic ones with laser excitation, such as optical absorption \cite{liu14,tran17}, Raman \cite{tran17,levs11,liu13} and Rayleigh spectroscopies \cite{liu13,zhao14,zhao20}.

The spectroscopic signal from nanotubes is normally observed for laser excitation close to their optical transitions. Therefore, the optical characterization of the nanotubes requires the precise theoretical modeling of the optical properties of the nanotubes, and, in particular, deriving their optical transitions. In the case of single-walled carbon nanotubes (SWNTs), the presence of helical symmetry allows for the reduction of the computational efforts for calculation of the optical transitions \cite{popo04}. In the approximation of neglecting the interlayer interaction, the optical properties of the DWNTs are determined solely by those of the layers and, in particular, the optical transitions of a DWNT are those of the two layers. While this approximation can be used for quick assignment of the optical spectra to DWNTs with specific layers, it is often observed that the optical transitions of DWNTs are shifted with respect to the corresponding ones of the layers, the deviations being attributed to the interlayer interaction \cite{levs11}. Since the observed shifts can be as large as several tens of meV, the mentioned approximation can yield incorrect identification of the layers. 

The estimation of the effect of the interlayer interaction on the optical transitions of the DWNTs has turned out to be a difficult computational problem because of the low symmetry of these structures. The shift of the optical transitions has been calculated by perturbation theory for a few tens of DWNTs \cite{liu14}. Using the effective theory and atomic structure mapping, it has been revealed that the electronic structure of the DWNTs can undergo a wide range of interlayer-interaction induced changes \cite{kosh15}. Recently, a number of optical resonances have been observed in the Rayleigh spectra of individual (free-standing) DWNTs, some of which cannot be connected to transitions of the layers \cite{zhao20}. The prediction of the optical resonances in the optical spectra of DWNTs is crucial for their structural characterization. As far as we are aware, such investigation within a realistic non-perturbative microscopic approach has not been reported yet.

Here, we study the effect of the interlayer interaction on the optical transitions of a particular DWNT by calculating the electronic density of states (DOS) and absorption coefficient using the recursion method with non-orthogonal tight-binding (NTB) basis functions. For many years, the recursion method has been the method of choice for calculation of transport properties of layered carbon structures within the orthogonal tight-binding approach (e.g., \cite{lamb00}) but, to our knowledge, it has not been used so far for realistic prediction of the optical properties of such structures.
 
The paper is organized as follows. The theoretical details are given in Sec. II. The obtained results are presented in Sec. III and discussed in Sec. IV. The paper ends up with conclusions, Sec. V.

\section{Theoretical background}

The wave equation, describing the quantum-mechanical systems, is often cast in the form of a matrix eigenvalue problem. Solving the problem for disordered systems can be hindered by the very large dimensions of the involved matrices. A powerful method for diagonalizing large sparse symmetric matrices has been proposed by Lanczos \cite{lanc50}. In this method, one selects an initial vector, constructs Krylov subspaces by matrix-vector products, and performs a three-term recurrence to finally obtain a new matrix in a tridiagonal form. This algorithm is faster than the direct diagonalization methods only for sparse matrices, for which the multiplication of the matrix and the vector can scale linearly with the dimension of the matrix.

The Lanczos method finds a particular application to electronic structure calculations for non-crystalline solids, where the algorithm is generally referred to as the Lanczos - Haydock method or the {\it recursion method} \cite{hayd75}. In the case of short-range interactions, it is advantageous to use the tight-binding approximation, where the Hamiltonian, arising from the expansion of the wavefunction as a linear combination of atomic orbitals, is obtained in the form of a sparse matrix. The latter is tridiagonalized by the Lanczos method and the corresponding real-space Green's function is derived. Then, the electronic DOS, electron density, total number of electrons, etc., are expressed through the real-space Green's function. 

The theoretical details on the calculation of the electronic DOS and the absorption coefficient, as well as on the recursion method, are provided in Appendices A, B, and C, respectively. 

\section{Results}

The calculations of the DOS and absorption coefficient are performed with NTB parameters taken over from density-functional theory (DFT) studies on carbon dimers. Two separate sets of parameters are used for description of the intralayer\cite{popo04} and the interlayer \cite{popo14} interactions. The Hamiltonian and overlap matrix elements between the $s$, $p_x$, $p_y$, and $p_z$ orbitals for the four valence electrons of the carbon atoms are obtained by substituting the NTB parameters into the Slater - Koster relations. Due to the localized nature of the atomic orbitals, the matrix elements  are nonzero only for atomic separations up to several~\AA. As a result, the Hamiltonian and overlap matrices are essentially sparse with normally up to $200$ nonzero elements in each row and column. Previously, the NTB parameters have been used for the successful prediction of the electronic structure and optical absorption of a large number of SWNTs \cite{mich09} and twisted bilayer graphene \cite{popo17}. For better agreement with experiment, the transition energies of the SWNTs have been rigidly upshifted by $0.44$ eV for transitions $S_{33}$, $S_{44}$... . This correction is implied everywhere below. 

\begin{figure}[tbph]
\includegraphics[width=70mm]{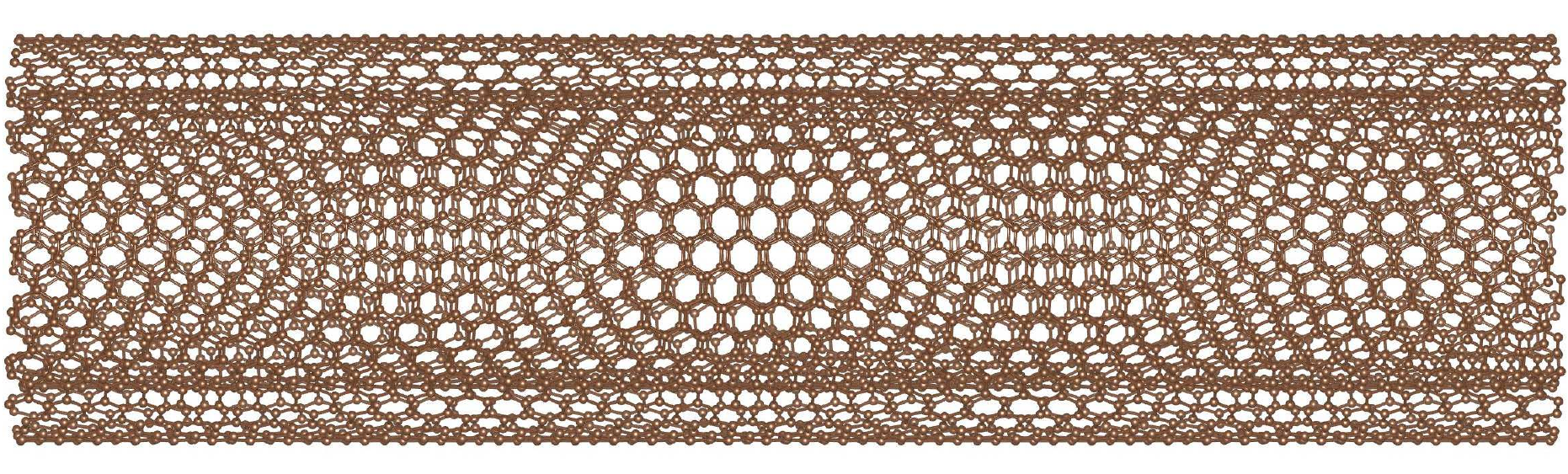}
\caption{Schematic of the atomic structure of a $100$~\AA-long piece the DWNT $(15,13)@(21,17)$. The radius of the inner (outer) layer is $9.51$~\AA ($12.91$~\AA) and the interlayer separation is $3.40$~\AA. }
\end{figure}

Here, the proposed computational scheme is applied to the case of the DWNT $(15,13)@(21,17)$ (Fig. 1) with recently reported  experimental Rayleigh spectrum \cite{zhao20}. In the usual DWNT notation \cite{pfei08}, $(15,13)$ are the chiral indices of the inner layer and $(21,17)$ are the chiral indices of the outer layer. A long piece of the DWNT of length $L$ and number of orbitals $N$ (number of atoms $N/4$) is considered. The atomic structure of the DWNT is relaxed as in Ref. \cite{popo18}. The recursion procedure is terminated at the $n$th recursion level. In the calculations of the DOS and absorption coefficient, different values of $L$, $N$, and $n$ are used, which are sufficient for deriving converged results in the energy interval between $1.1$ and $2.9$ eV. The large size of the considered piece of the DWNT ensures negligible influence of the edge states on the DOS and absorption coefficient.

\begin{figure}[tbph]
\includegraphics[width=70mm]{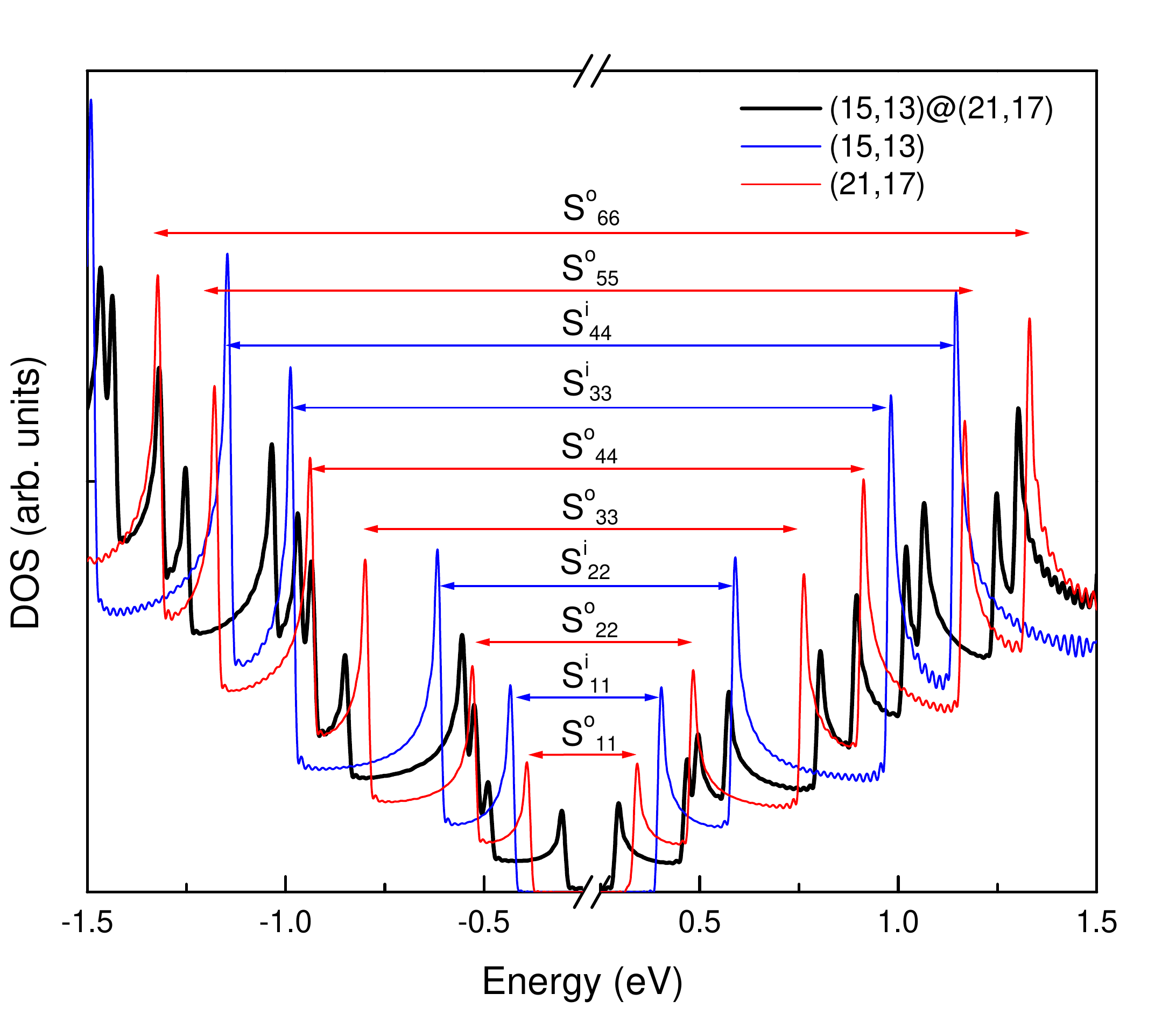}
\caption{The DOS of the DWNT $(15,13)@(21,17)$ (black line) in comparison with the DOS of the non-interacting inner layer $(15,13)$ (blue line) and outer layer $(21,17)$ (red line). The horizontal arrows between the mirror spikes mark the optical transitions for the non-interacting layers, denoted by $S^i$ and $S^o$ for the inner and outer layer, respectively. }
\end{figure}

Figure 2 presents the results for the DOS of the DWNT, obtained with $L=2 000$~\AA, $N=430 000$, and $n=5 000$, in comparison with the DOS of the non-interacting layers. According to the selection rules for optical transitions for light polarization along the axis of the SWNT, the optical transitions take place between mirror spikes of the DOS. The number of obtained eigenenergies by the recursion method is of the same order as $n$, and is sufficiently large to allow determining the optical transitions with accuracy of $0.01$ eV. The so-derived optical transitions of the non-interacting inner ({\it i}) and outer ({\it o}) layers, denoted by $S^i$ and $S^o$, respectively, correspond within $0.01$ eV to the already derived within the NTB model by solving Eq. (\ref{a30}) with direct diagonalization \cite{popo04}. We note that the derivation of such large number of eigenenergies by the DFT approach, either by directly solving the DFT equations, or by using the recursion method, is computationally expensive and has not been reported for DWNTs so far.

It is clear from Fig. 2 that the DOS of the DWNT undergoes significant changes, most of the spikes being red or blue shifted with respect to those of the non-interacting layers. Since the number of spikes of the DOS of the DWNT corresponds to that of the layers, it is tempting to derive the optical transitions of the DWNT as the separation between the mirror spikes of the DOS and adopt the same notation as for the layers. The so-derived optical transitions of the DWNT are given in Table 1 in comparison with those for the non-interacting layers and the experimentally measured ones. It is seen from Table 1 that the transitions of the DWNT can have large shifts with respect to those of the non-interacting layers. 

\begin{table}[width=.9\linewidth,cols=7,pos=h]
\caption{\label{tab:table1}
Optical transition energies (in eV) of the DWNT (first line) and the non-interacting layers (second line), derived from the separation between the mirror spikes of DOS, together with the shift of the former with respect to the latter (fourth line). Available experimental values are provided for comparison \cite{liu12} (third line).
}
\begin{tabular*}{\tblwidth}{@{} LRRRRRRR@{} }
\toprule
& $S^i_{22}$ & $S^o_{33}$ & $S^o_{44}$ & $S^i_{33}$ & $S^i_{44}$ & $S^o_{55}$ & $S^o_{66}$ \\
\midrule
DWNT & $1.11$ & $1.63$ & $1.81$ & $1.97$ & $2.09$ & $2.48$ & $2.60$ \\
layer   & $1.19$ & $1.54$ & $1.83$ & $1.95$ & $2.27$ & $2.33$ & $2.63$ \\
layer \cite{liu12} & $-$ & $1.44$ & $1.82$ & $1.94$ & $2.28$ & $2.34$ & $-$ \\
shift    & $-0.08$ & $0.09$ & $-0.02$ & $0.02$ & $-0.18$ & $0.15$ & $-0.03$ \\
\bottomrule
\end{tabular*}
\end{table}

\begin{figure}[tbph]
\includegraphics[width=70mm]{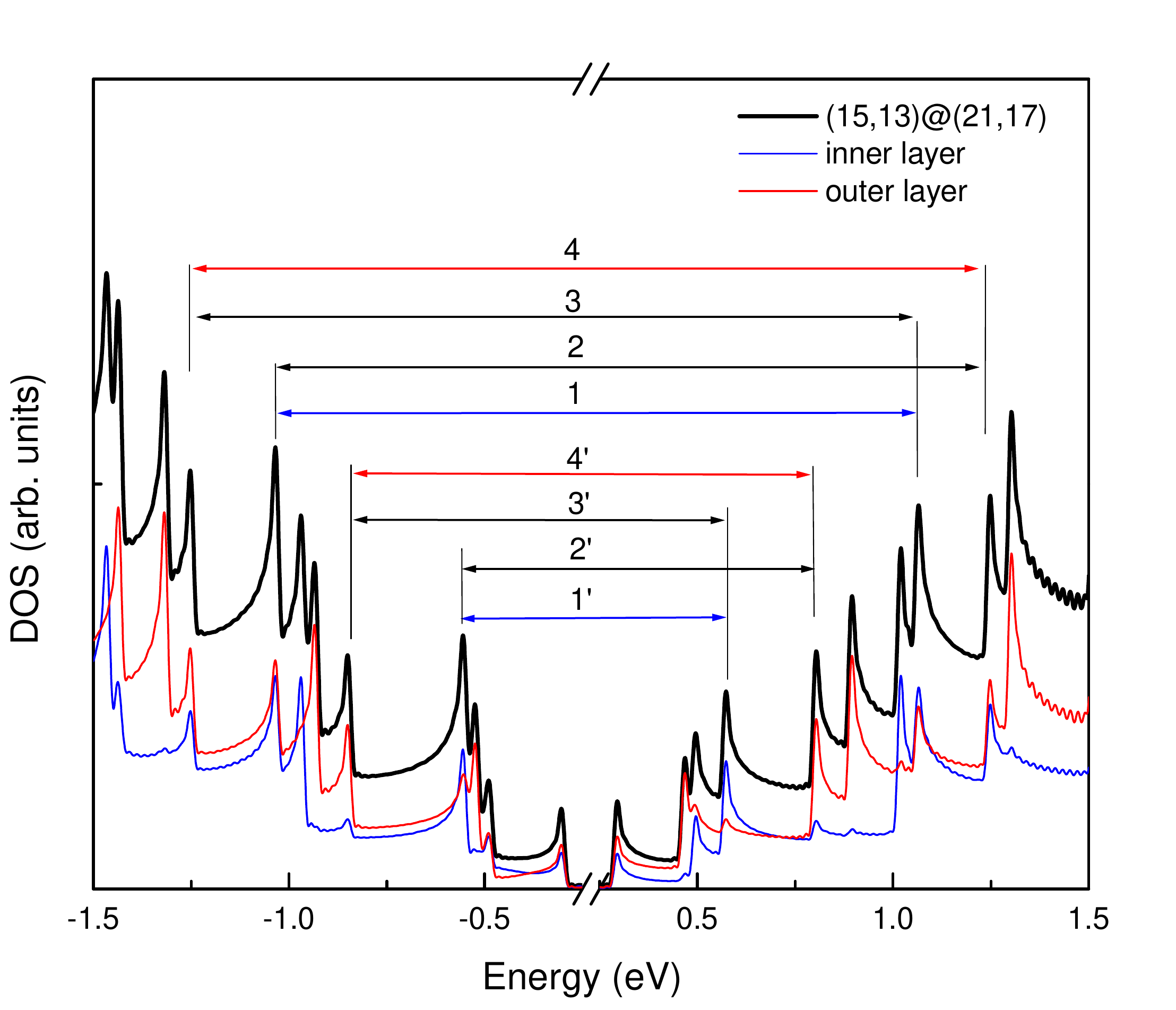}
\caption{The DOS of the DWNT $(15,13)@(21,17)$ (black line) in comparison with the contributions of the inner layer $(15,13)$ (blue line) and outer layer $(21,17)$ (red line). The graph shows strong mixing of electronic states of the two layers close to energies, marked by vertical lines. The red and blue horizontal arrows show transitions between mirror spikes corresponding to specific pairs of transitions of the layers, while the black horizontal arrows show cross-band transitions, induced by the mixing of the electronic states.}
\end{figure}

Such a simplified approach to the derivation of the optical transitions of the DWNT rules out the possibility of appearance of additional optical transitions. On the other hand, the electronic structure of the DWNT is significantly modified with respect to that of the layers due to the interlayer interaction and, therefore, new transitions cannot be excluded {\it a priori}. For elucidating this problem, we plot in Fig. 3 the DOS of the DWNT in comparison with the contributions of the layers. It is seen in Fig. 3 that most of the spikes of DOS of the DWNT can be connected to one of the layers. However, the four spikes, marked by vertical lines and connected by arrows $1$ to $4$, have non-negligible contribution from both layers, which can be interpreted as a significant mixing of the electronic states of the two layers. Thus, four transitions can be expected to appear because of  the mixing of the mentioned states: two transitions, $1$ and $4$, between mirror spikes, corresponding to transitions  $S^i_{44}$ and $S^o_{55}$ of the non-interacting layers, and two cross-band transitions, $2$ and $3$, with energies $2.26$ eV and $2.31$ eV, respectively. Similarly, large mixing is evident for the spikes, connected by arrows $1^\prime$ to $4^\prime$. Therefore, four transitions can be expected between these spikes: two transitions, $1^\prime$ and $4^\prime$, between mirror spikes, corresponding to transitions $S^i_{22}$ and $S^o_{33}$ of the non-interacting layers, and two cross-band transitions, $2^\prime$ and $3^\prime$, with energies $1.34$ eV and $1.40$ eV.

\begin{figure}[tbph]
\includegraphics[width=70mm]{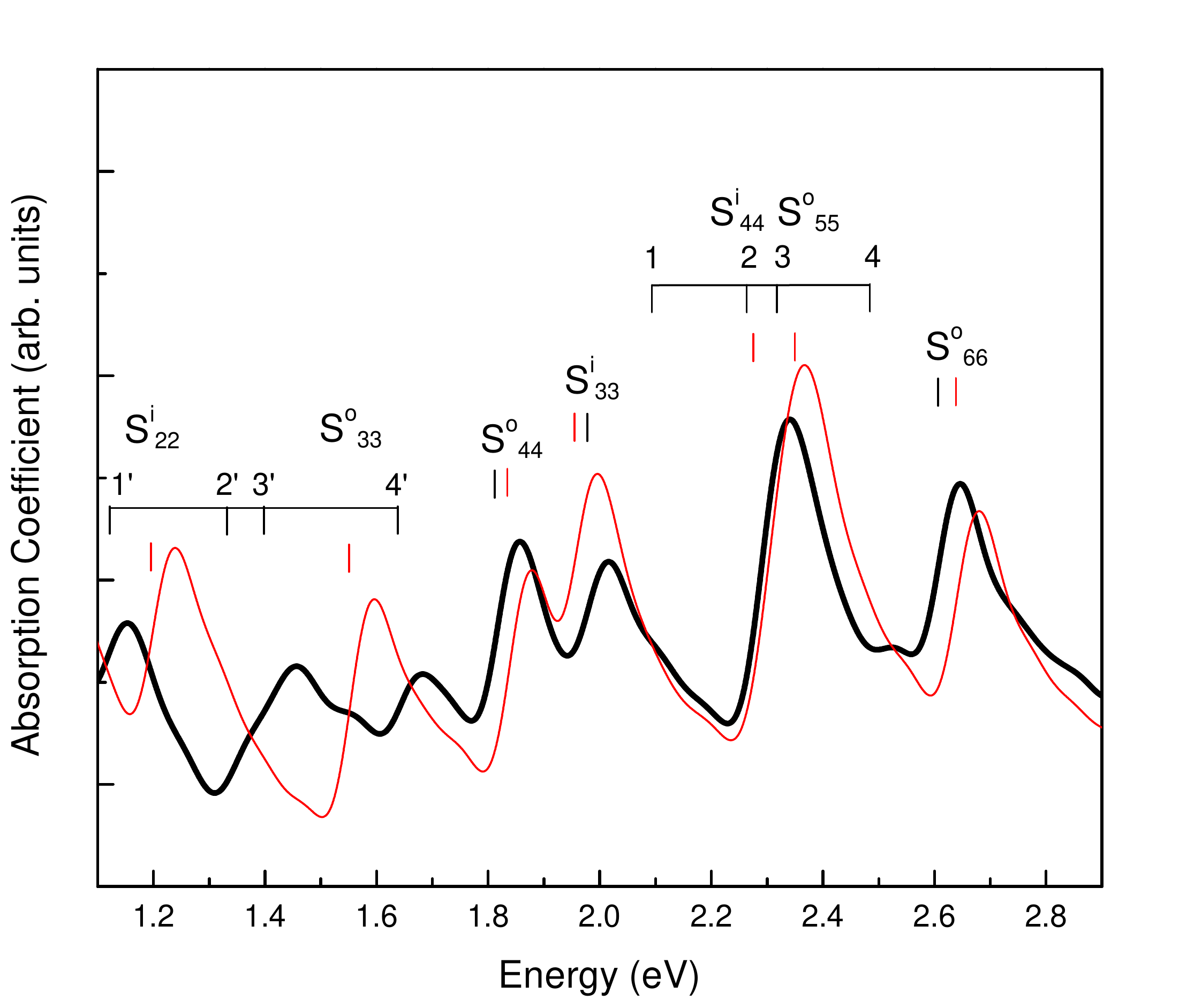}
\caption{The absorption coefficient of the DWNT $(15,13)@(21,17)$ (thick black line) in comparison with that for a DWNT without interlayer interaction (thin red line). The red vertical lines mark the transitions of the layers. The black vertical line mark the transitions of the DWNT. The numbers $1$ to $4$ and $1^\prime$ to $4^\prime$ mark the transitions, shown in Fig. 3.}
\end{figure}

Figure 4 shows the calculated absorption coefficient of the DWNT, obtained with $L=400$~\AA, $N=86 000$, and $n=1 000$, in comparison with that of the non-interacting layers.  It can be expected that the major changes in the electronic structure will manifest themselves in the absorption coefficient. It is seen in Fig. 4, that the spectrum of the DWNT differs significantly from that of the non-interacting layers, having a larger number of features. 

As argued above, four transitions, $1$ to $4$, and another four transitions, $1^\prime$ to $4^\prime$, in Fig. 3, can be expected to give rise to features in the absorption spectrum. It is seen in Fig. 4 that at the energies of transitions $1$ and $4$ there are only a tiny kink and a small bump, while at the energies of transitions $2$ and $3$ there is a high peak. Therefore, the cross-band transitions give major contribution to the absorption, contrary to the simplified approach, within which this peak is ascribed to transitions between mirror spikes of DOS. Transitions $1^\prime$ and $4^\prime$ give rise to smaller peaks than for the non-interacting layers, while transitions $2^\prime$ and $3^\prime$ contribute to a wide peak. As a whole, all transitions $1^\prime$ to $4^\prime$ give rise to peaks of comparable height. Other smaller features are also present in the absorption. For example, the kink at $\approx 1.5$ eV is due to cross-band transitions between states, related to transitions $S^i_{22}$ and $S^o_{44}$ of the non-interacting layers. For gaining physical insight into the obtained results, in the next section, we provide a simple description of the mixing of electronic states of the layers and arguments for the appearance of new optical transitions.

\section{Discussion}

\subsection{Transitions $1$ to $4$}

Transitions $1$ to $4$ of the DWNT are related to transitions $S^i_{44}$ and $S^o_{55}$, which take place  between almost overlapping spikes of the non-interacting layers. The former can be derived from a simple model of the DWNT, in which the inner and outer layers are replaced by two identical systems in states with wavefunctions $\psi^i$ and $\psi^o$, and equal energies $E^i = E^o \equiv E$.  The lifting of the degeneracy upon switching-on the interaction between the systems can be studied by the time-independent quantum-mechanical perturbation theory for degenerate energy levels (e.g., \cite{bohm89}). For finding the perturbed wavefunctions and energies, we consider the interaction between the systems as a perturbation, described by the operator $\hat V$ and choose the zeroth-order wavefunction $\psi$ as the linear combination $\psi =\alpha\psi^i +\beta\psi^o$. Substituting $\psi$ in the Schroedinger equation for the coupled systems, we obtain a system of two linear equations with solutions
\begin{align}
E_k=&E+\frac{V^{ii}+V^{oo}}{2}
 \pm \left(\frac{\left(V^{ii}-V^{oo}\right)^2}{4}+|V^{io}|^2\right)^{1/2},\label{a50}\\
\frac{\alpha_k}{\beta_k}=&\frac{V^{io}}{E_k-E-V^{ii}}.
\label{a51}
\end{align}
Here, $k=1,2$; $V^{\mu\nu}$, $\mu,\nu=i,o$, are the matrix elements of $\hat V$ between $\psi^i$ and $\psi^o$; $V^{io}=V^{oi*}$.

Assuming for simplicity that $V^{ii}=V^{oo}$ and $V^{io}=V^{oi}$, and using Eqs. (\ref{a50}), (\ref{a51}), we obtain the wavefunctions of the higher and lower-energy states as
\begin{align}
\psi_1=&(\psi^i + \psi^o)/\sqrt{2},\label{a52}\\
\psi_2=&(\psi^i - \psi^o)/\sqrt{2}.\label{a53}
\end{align} 
 
Similarly, for two pairs of states of the non-interacting systems: a pair of occupied ($v$) states $\psi^i_v$ and $\psi^o_v$ with equal energies $E^i_v = E^o_v \equiv E_v$, and a pair of unoccupied ($c$) states $\psi^i_c$ and $\psi^o_c$ with equal energies $E^i_c = E^o_c \equiv E_c$ ($E_c >E_v$), we obtain the wavefunctions of the coupled systems in order of decreasing energy as
\begin{align}
\psi_{c1}=(\psi^i_c + \psi^o_c)/\sqrt{2},\label{a54}\\
\psi_{c2}=(\psi^i_c - \psi^o_c)/\sqrt{2},\label{a55}\\
\psi_{v1}=(\psi^i_v+ \psi^o_v)/\sqrt{2},\label{a56}\\
\psi_{v2}=(\psi^i_v- \psi^o_v)/\sqrt{2}.\label{a57}
\end{align} 

If transitions $E^i_v \to E^i_c$ and $E^o_v \to E^o_c$ are allowed in the non-interacting layers, in the DWNT, they will be doubled to four transitions, which can be labeled $1$ to $4$ in order of increasing energy. Then, transitions $1$ and $4$ will be mirror ones and $2$ and $3$ will be cross-band ones.

The height of the absorption peaks is determined by the squared modulus of the matrix element of the momentum $\hat p$ between the wavefunctions, Eqs. (\ref{a54}) - (\ref{a57}). This matrix element can be expressed via the matrix element $p^{\mu\nu}$ of $\hat p$ between the wavefunctions $\psi^{\mu}_c$ and $\psi^{\nu}_v$, $\mu , \nu=i,o$. After some algebra and taking into account that, for $\mu\ne \nu$, the matrix element $p^{\mu\nu}$ is small and can be neglected, we obtain 
\begin{align}
|p_{1}|^2=&|p_{4}|^2=|p^{ii}-p^{oo}|^2/2,\label{a58}\\
|p_{2}|^2=&|p_{3}|^2=|p^{ii}+p^{oo}|^2/2,\label{a59}
\end{align}
where the lower index of matrix element labels the transition.

Therefore, the two absorption peaks due to the mirror transitions (cross-band transitions) will be of the same height. It will depend on the matrix elements $p^{ii}$ and $p^{oo}$, and may differ significantly between the two groups of transitions. 

The presented arguments can be applied to the case of nearly-degenerate states for $\alpha =|(E^i - E^o)/V^{io}| \ll 1$. Then, Eqs. (\ref{a58}), (\ref{a59}), will be corrected by additive terms of second and higher order of smallness with regard to $\alpha$.

The simplified picture of mixing of the states and doubling the transitions is in accord with the full calculations from the previous Section. The obtained here significant mixing of the states, described by the wavefunctions Eqs. (\ref{a54}) - (\ref{a57}), is clearly seen in Fig. 3. Equations (\ref{a58}) and (\ref{a59}) with $p^{ii}= p^{oo}$ predict a single peak for the cross-band transitions $2$ and $3$ but no peaks for the mirror transitions $1$ and $4$, which corresponds to the calculated spectrum in Fig. 4. 

We note that the size of the splitting of the degenerate level is determined by the matrix element of the interlayer interaction, $V^{io}$. The selection rules for the latter depend on the symmetry properties of the wavefunctions, which are connected to the atomic structure of the layers, defined by their chiral indices. Therefore, the interaction-induced modifications of the DOS and absorption coefficient are essentially chirality-dependent. In particular, different modifications can be expected for DWNTs with similar radii but different chiralities.

\subsection{Transitions $1^\prime$ to $4^\prime$}

Transitions $1^\prime$ to $4^\prime$ of the DWNT are related to transitions $S^i_{22}$ and $S^o_{33}$  of the non-interacting layers with energies, differing by $0.35$ eV. We adopt the same simplified model of the DWNT as above but with different energies $E^i$ and $E^o$. The modification of the wavefunctions and energies upon switching-on the interaction $\hat V$ can be described by the time-independent quantum-mechanical perturbation theory for nondegenerate energy levels (e.g., \cite{bohm89}). It can be demonstrated that the perturbation shifts the two energies, and the modified wavefunctions are given, up to first-order in the perturbation, by
\begin{align}
\psi_1=&(\psi^i -\alpha \psi^o)/\sqrt{1+\alpha ^2},  \label{a72}\\
\psi_2=&(\alpha \psi^i + \psi^o)/\sqrt{1+\alpha ^2}.\label{a73}
\end{align} 
Here, the indices $1$ and $2$ refer to the first and second systems, corresponding to the inner and outer layer, respectively; $\alpha=V^{oi}/(E^o - E^i)$. The condition for applicability of the perturbation theory is $|\alpha| \ll 1$.

Considering  two pairs of states of the non-interacting systems: a pair of occupied ($v$) states $\psi^i_v$ and $\psi^o_v$ with  energies $E^i_v > E^o_v$, and a pair of unoccupied ($c$) states $\psi^i_c$ and $\psi^o_c$ with energies $E^i_c < E^o_c$, we find the modified wavefunctions as
\begin{align}
\psi_{c1}=&(\psi_c^i -\alpha_c \psi_c^o)/\sqrt{1+\alpha_c ^2},  \label{a74}\\
\psi_{c2}=&(\alpha_c \psi_c^i + \psi_c^o)/\sqrt{1+\alpha_c ^2},\label{a75}\\
\psi_{v1}=&(\psi_v^i -\alpha_v \psi_v^o)/\sqrt{1+\alpha_v ^2},  \label{a76}\\
\psi_{v2}=&(\alpha_v \psi_v^i + \psi_v^o)/\sqrt{1+\alpha_v ^2},\label{a77}
\end{align} 
where $\alpha_\sigma=V_\sigma^{oi}/(E_\sigma^o - E_\sigma^i)$, $\sigma = c,v$.

If transitions $E^i_v \to E^i_c$ and $E^o_v \to E^o_c$ are allowed in the non-interacting layers,  in the DWNT, they will be doubled to four transitions, which can be labeled $1^\prime$ to $4^\prime$ in order of increasing energy. The transitions $1^\prime$ and $4^\prime$ will be mirror ones and $2^\prime$ and $3^\prime$ will be cross-band ones.

Finally, the height of the absorption peaks is determined by the squared matrix element of the momentum between the wavefunctions Eqs. (\ref{a74}) - (\ref{a77}), which can be expressed by the matrix element $p^{\mu\nu}$, $\mu, \nu=i,o$, 
\begin{align}
|p_{1}|^2= &| p^{ii}-\alpha^2p^{oo}|^2/(1+\alpha^2),\label{a78}\\
|p_{4}|^2= &|\alpha^2p^{ii}- p^{oo}|^2/(1+\alpha^2),\label{a79}\\
|p_{2}|^2=&|p_{3}|^2= \alpha^2|p^{ii}+p^{oo}|^2/(1+\alpha^2).\label{a80}
\end{align}
Here, it is assumed for simplicity that  $|\alpha_c| =|\alpha_v| \equiv \alpha$. 

It follows from Eqs. (\ref{a78}) and (\ref{a79}) that the mirror transitions $1^\prime$ and $4^\prime$ will give rise to peaks, which are smaller than those for the non-interacting systems. Equation (\ref{a80}) shows that the cross-band transitions $2^\prime$ and $3^\prime$ will have equal contributions to the absorption. With decreasing $\alpha$, the height of the cross-band peak(s) will decrease and disappear in the limit $\alpha \to 0$.

Equations  (\ref{a78}) - (\ref{a80}) with $\alpha \approx 0.3$ predict peaks of comparable height in agreement with the full calculations of the previous Section, presented in Fig. 4.  

The interaction-induced mixing of the states of the two layers, as well as shifting of the corresponding energies, is determined by the matrix element of the interlayer interaction, $V^{io}$, which depends on the symmetry of the wavefunctions of the states. Therefore, the changes of the DOS and absorption coefficient will be chirality-dependent.

\subsection{Comparison to experiment}

In a recent paper \cite{zhao20}, the observed Rayleigh resonances of the considered DWNT have been assigned to shifted optical transitions of the non-interacting layers. Based on the calculated optical absorption, we make a different conclusion about the origin of some of the optical resonances in the experimental spectrum. Namely, our study provides a theoretical evidence that, apart from shifts of the transitions, new transitions also appear. In particular, the resonance around $2.3$ eV is predicted here to arise from such new transitions, rather than from shifted transitions of the layers. 

\section{Conclusions}

We have studied the effect of the interlayer interaction on the electronic structure and optical absorption of the DWNT $(15,13)@(21,17)$  using the recursion method with NTB basis functions. This method has the advantage to describe the effect of the interlayer interaction within the quantum-mechanical picture without resorting to the perturbation theory. The recursion method is used with ab-initio derived NTB parameters, which yield realistic prediction of the optical transitions of layered carbon structures. 

In particular, the calculations reveal that the interlayer interaction can give rise to major changes of the electronic structure, such as mixing of the states of the two layers and shift of their energies, which may be accompanied by appearance of new optical transitions. Strong mixing of the electronic states can be expected if the two non-interacting layers have close optical transitions and there is a strong interaction between the electronic states of the layers. Significant mixing of the electronic states can be expected even for relatively large difference of the transition energies of the non-interacting layers, which can result in absorption peaks of the cross-transitions, comparable to those of the mirror transitions. We underline that the adopted NTB description of the interlayer coupling yields chirality-dependent modifications of the DOS and absorption coefficient of DWNTs. 

The predicted absorption coefficient of the DWNT is in excellent agreement with the available experimental data. The presented computational approach can be used for the realistic prediction of the optical absorption of synthesized DWNTs and twisted few-layer graphene for the needs of their structural characterization.

\section*{Acknowledgments}

VNP acknowledges financial support from the National Science Fund of Bulgaria under grant KP-06-N38/10-06.12.2019. VNP thanks Profs. M. Paillet and J.-L. Sauvajol for fruitful discussions.

\appendix

\section{The electronic density of states}
The quantum-mechanical description of an atomic system is usually based on the Schroedinger time-independent wave equation  \cite{gros00}
\begin{equation}
\hat{H}\psi_{\lambda}({\bf r}) = E_{\lambda}\psi_{\lambda}({\bf r}),
\label{a1}
\end{equation}
where $\hat{H}$ is the spin-independent Hamiltonian of the system, $\psi_{\lambda}({\bf r})$ is the wavefunction, $E_{\lambda}$ is the energy, and the index $\lambda$ enumerates the solutions of the wave equation.
In the NTB approach, the wavefunction is expanded as a linear combination of atomic orbitals
\begin{equation}
\psi_{\lambda}({\bf r}) = \sum_{\alpha}C_{\alpha}^{\lambda}\varphi_{\alpha}({\bf r}),
\label{a2}
\end{equation}
where $C_{\alpha}^{\lambda}$ are expansion coefficients, $\varphi_{\alpha}$ are atomic orbitals, and the index $\alpha$ runs over the atomic orbitals in the solid: $\alpha=1,2,...,N$.
The substitution of Eq. (\ref{a2}) in Eq. (\ref{a1}) results in the matrix eigenvalue equation 
\begin{equation}
\sum_{\beta} (H_{\alpha\beta} - E_{\lambda}S_{\alpha\beta}) C_{\beta}^{\lambda} = 0.
\label{a30}
\end{equation}
Here 
\begin{equation}
H_{\alpha\beta} = \int \varphi_{\alpha}^{\ast}({\bf r})\hat{H}\varphi_{\beta}({\bf r}) d{\bf r} 
\label{a31}
\end{equation} 
are the Hamiltonian matrix elements with respect to the atomic orbitals and 
\begin{equation}
S_{\alpha\beta} = \int \varphi_{\alpha}^{\ast}({\bf r})\varphi_{\beta}({\bf r}) d{\bf r} 
\label{a32}
\end{equation}
are the overlap matrix elements, arising from the non-orthogonality of orbitals of different atoms.
From the normalization condition for $\psi_{\lambda}$ 
\begin{equation}
\int \psi_{\lambda}^{\ast}({\bf r})\psi_{\lambda^{\prime}}({\bf r})d{\bf r} 
= \delta_{\lambda\lambda^{\prime}},
\label{a5}
\end{equation}
one obtains
\begin{equation}
\sum_{\alpha\beta} C_{\alpha}^{\lambda^{\ast}}S_{\alpha\beta}C_{\beta}^{\lambda^{\prime}} = \delta_{\lambda\lambda^{\prime}},
\label{a33}
\end{equation}
where $\delta_{\lambda\lambda^{\prime}}$ is the Kronecker delta.

The electron density is given by 
\begin{align}
\rho({\bf{r}}) &= 2 \sum_{\lambda}^{occ} \mid \psi_{\lambda}({\bf{r}}) \mid ^2 \\
&=  2 \int_{-\infty}^{E_F} \sum_{\lambda}\delta (E-E_{\lambda}) \mid \psi_{\lambda}({\bf{r}}) \mid ^2 dE\\
&\equiv \sum_{\alpha\beta}  {\rho}_{\beta\alpha} \varphi_{\alpha}^{\ast}({\bf{r}})\varphi_{\beta}({\bf{r}}),
\label{a34}
\end{align}
where the summation over $\lambda$ is carried out over all occupied states up to the Fermi energy $E_F$ and the factor $2$ accounts for the spin degeneracy; $\delta (E-E_{\lambda})$ is the Dirac delta function. In the last line of Eq. (\ref{a34}), Eq. (\ref{a2}) is used and the following notation is introduced
\begin{equation}
{\rho}_{\beta\alpha} = \int_{-\infty}^{E_F} {\rho}_{\beta\alpha}(E) dE,
\label{a35}
\end{equation}
where
\begin{equation}
{\rho}_{\beta\alpha}(E) = 2  \sum_{\lambda}\delta (E-E_{\lambda}) C_{\alpha}^{\lambda\ast}C_{\beta}^{\lambda}.
\label{a36}
\end{equation}

The DOS $\rho(E) = 2\sum_{\lambda}\delta (E-E_{\lambda})$ can be written as
\begin{equation}
\rho(E)=\sum_{\alpha\beta}{\rho}_{\alpha\beta}(E)S_{\beta\alpha}.
\label{a37}
\end{equation}

The matrix ${\rho}_{\beta\alpha}(E)$ can be connected to the imaginary part of the Green's function $G_{\beta\alpha}({\tilde{E}})$ (${\tilde{E}}=E+i\eta$, $\eta \to 0^{+}$) 
\begin{equation}
G_{\beta\alpha}({\tilde{E}}) =\sum_{\lambda} ({\tilde{E}} - E_{\lambda})^{-1}C_{\alpha}^{\lambda^{\ast}}C_{\beta}^{\lambda},
\label{a41}
\end{equation}
namely,
\begin{equation}
\rho_{\beta\alpha}(E) =-(2/\pi) \Im G_{\beta\alpha}({\tilde{E}}).
\label{a43}
\end{equation}
Therefore, the knowledge of the Green's function allows for the calculation of the DOS, Eq. (\ref{a37}).

\section{The optical absorption coefficient}
The one-photon optical absorption in nanotubes is usually observed for light polarization along the nanotube. The optical absorption coefficient can be expressed via the imaginary part of the frequency-dependent dielectric function $\epsilon_2(\omega)$, given by \cite{gros00}
\begin{equation}
\epsilon_2(\omega) \propto \frac{1}{\omega^2}\sum_{\lambda}^{occ}\sum_{\lambda^{\prime}}^{unocc}|p_{\lambda^{\prime}\lambda}|^2\delta(E_{\lambda^{\prime}}-E_{\lambda}-\hbar\omega),
\label{a60}
\end{equation}
where $p_{\lambda^{\prime}\lambda}$ is the momentum matrix element
\begin{equation}
p_{\lambda^{\prime}\lambda} = \int \psi_{\lambda^{\prime}}^{\ast}({\bf r})\hat{p}\psi_{\lambda}({\bf r}) d{\bf r} 
\label{a61}
\end{equation}
and $\hat{p}$ is the component of the momentum operator along the nanotube; $\lambda$ runs over the occupied states and $\lambda^{\prime}$ runs over the unoccupied states. 

Here, using Eq. (\ref{a2}), we cast the matrix element $p_{\lambda^{\prime}\lambda}$ in the form
\begin{equation}
p_{\lambda^{\prime}\lambda}=\sum_{\alpha\beta}C_{\alpha}^{\lambda^{\prime}\ast}p_{\alpha\beta}C_{\beta}^{\lambda},
\label{a62}
\end{equation}
where 
\begin{equation}
p_{\alpha\beta} = \int \varphi_{\alpha}^{\ast}({\bf r})\hat{p}\varphi_{\beta}({\bf r}) d{\bf r}. 
\label{a63}
\end{equation} 
In view of $p_{\lambda^{\prime}\lambda}^{\ast}=p_{\lambda\lambda^{\prime}}$, we get 
\begin{equation}
|p_{\lambda^{\prime}\lambda}|^2=p_{\lambda^{\prime}\lambda}^{\ast}p_{\lambda^{\prime}\lambda}=\sum_{\alpha\beta\gamma\delta}C_{\alpha}^{\lambda^{\prime}}C_{\gamma}^{\lambda^{\prime}\ast}p_{\gamma\delta}C_{\delta}^{\lambda}C_{\beta}^{\lambda\ast}p_{\beta\alpha}.
\label{a64}
\end{equation}
Introducing $\rho_{\alpha\beta}(E)$, Eq. (\ref{a36}), $\epsilon_2(\omega)$ becomes
\begin{align}
\epsilon_2(\omega) \propto &
\frac{1}{\omega^2}\int dE\int dE^{\prime}\times \nonumber\\
  &\times \sum_{\alpha\beta} p_{\alpha\beta}(E) p_{\beta\alpha}(E^{\prime})  {\delta (E^{\prime}-E-\hbar\omega)},
\label{a65}
\end{align}
where $p_{\alpha\beta}(E)=\sum_{\gamma}\rho_{\alpha\gamma}(E) p_{\gamma\beta}$. The integration is performed over the occupied states with energy $E$ and the unoccupied states with energy $E^{\prime}$. It is clear that the calculation of $\epsilon_2(\omega)$ requires the knowledge of $\rho_{\alpha\beta}(E)$, but not of the wavefunction.

Finally, the absorption coefficient $\alpha(\omega)$ can be evaluated approximately as
\begin{equation}
\alpha(\omega) \propto \omega\epsilon_2(\omega).
\label{a66}
\end{equation}

\section{The recursion method}
For the calculation of ${\bf G}({\tilde{E}})$, the Hamiltonian ${\bf H}$ is tridiagonalized by the modified three-term recurrence\cite{jone84}
\begin{equation}
b_{i+1}{\bf u}_{i+1} = ({\bf H}^{\prime}-a_{i}{\bf I}) {\bf u}_{i} - b_{i}{\bf u}_{i-1}.
\label{a44}
\end{equation}
Here, ${\bf H}^{\prime} ={\bf S}^{-1}{\bf H}$ is an $N\times N$ matrix, ${\bf I}$ is an $N\times N$ unit matrix, ${\bf u}_{i}$ are $S$-orthonormal column-vectors of size $N$: ${\bf u}_{i}^{+}{\bf S}{\bf u}_{j}=\delta_{ij}$, $\delta_{ij}$ is the Kronecker delta; $a_{i}$ and $b_{i}$ ($i=1,2,...,n$, $n \le N$, $n$ is the number of recursion levels) are elements of the tridiagonal $n\times n$ matrix 
\begin{equation}
{\bf H}_{TD}=
 \begin{pmatrix}
  a_{1}  &  b_{2} & \cdots & 0 \\
  b_{2}  &  a_{2} & \cdots & 0 \\
  \vdots  & \vdots & \ddots & \vdots  \\
  0         & 0        & \cdots  & a_{n} 
 \end{pmatrix}
 \label{a45}
\end{equation}

During the recurrence procedure, partial reorthogonalization of ${\bf u}_{i}$ is performed to avoid the loss of orthogonality and appearance of ghost states due to the finite-precision arithmetic.

The recurrence, Eq. (\ref{a44}), can be written as the matrix equation
\begin{equation}
{\bf H}^{\prime}{\bf U} = {\bf U}{\bf H}_{TD},
\label{a46}
\end{equation}
where ${\bf U}$ is an $N\times n$ matrix consisting of the column-vectors ${\bf u}_{i}$. The orthonormality condition for  ${\bf u}_{i}$ can be written concisely as ${\bf U^{+}SU=I}$, where ${\bf I}$ is an $n\times n$ unit matrix and ${\bf U^{+}}$ is the Hermitian conjugate of ${\bf U}$.  

Next, using Eq. (\ref{a46}), the following relation between the Green's functions ${\bf G}=(\tilde{E}{\bf S}-{\bf H})^{-1}$ and ${\bf G}_{TD}=(\tilde{E}{\bf I}-{\bf H}_{TD})^{-1}$ is readily derived
\begin{equation}
{\bf U^{+}S G} = {\bf G}_{TD}{\bf U^{+}}.
\label{a47}
\end{equation}
The Green's function ${\bf G}_{TD}$ is expressed as a Jacobi continued fraction expansion, which is terminated at the $n$th recursion level and the square-root terminator is used for the remainder of the expansion.

Equation (\ref{a47}) with a starting vector $u_{1\beta}=\delta_{\alpha\beta}$ is reduced to $G_{\alpha\beta}^{\prime} = G_{TD,1\gamma}U^{+}_{\gamma\beta}$, where ${\bf G}^{\prime}={\bf S}{\bf G}$. After evaluating ${\bf G}^{\prime}$, the Green's function ${\bf G}$ is found as ${\bf G }={\bf S}^{-1} {\bf G}^{\prime}$.

The overlap matrix ${\bf S}$ can be inverted by the recursion method as well \cite{ozak01}. Indeed, the inverse of the overlap matrix can be written as ${\bf S}^{-1}= {\Re \bf R}(0)$, where ${\bf R}(\tilde{E}) = ({\bf{S}}-{\tilde{E}}{\bf I})^{-1}$ and ${\bf I}$ is an $N\times N$ unit matrix.
%

\bibliographystyle{model1-num-names}

\end{document}